\documentclass[aps,prl,twocolumn,showpacs,superscriptaddress]{revtex4}
\usepackage{graphicx}
\usepackage{hyperref}
\usepackage{natbib}
\begin{document}
\title{Quantum oscillations without quantum coherence}
\author{V. V. Dobrovitski}
\affiliation{Ames Laboratory, Iowa State University, Ames IA 50011, USA}
\author{H. A. De Raedt}
\affiliation{Institute for Theoretical Physics and Materials 
  Science Centre, University of Groningen, Nijenborgh 4, NL-9747 AG
  Groningen, The Netherlands}
\author{M. I. Katsnelson}
\affiliation{Institute of Metal Physics, Ekaterinburg 620219, Russia}
\author{B. N. Harmon}
\affiliation{Ames Laboratory, Iowa State University, Ames IA 50011, USA}
\begin{abstract}
We study numerically the damping of quantum oscillations and the
increase of entropy with time in model spin systems decohered by
a spin bath. In some experimentally relevant cases, the
oscillations of considerable amplitude can persist long after the
entropy has saturated near its maximum, i.e. when the system has
been decohered almost completely. Therefore, the pointer states
of the system demonstrate non-trivial dynamics. The oscillations
exhibit slow power-law decay, rather than exponential or
Gaussian, and may be observable in experiments.
\end{abstract}

\maketitle

For a quantum system prepared in a linear superposition of its
eigenstates, some observables can oscillate with time.
Interaction of the system with its environment leads to a decay
of the system's initial pure state into a mixture of several
"pointer states"; it causes an increase of the system's
entropy (decoherence) and damping of quantum oscillations
(dephasing) with time \cite{decgen,decgen1}. Both effects,
decoherence and dephasing, are often considered as equivalent
results of the mixed state of the system. But careful analysis
shows important differences \cite{decdeph,zeh1} originating
from the fact that
the same density matrix can describe both an
ensemble of similar systems, and a single decohered system. Thus,
e.g., dephasing can appear in an ensemble of pure, non-decohered
systems with slightly differing dynamics (this is an idealized
picture of $T_2$ processes in NMR). Decoherence and
dephasing are hard to distinguish in experiments which 
employ ensembles of quantum systems. However, recently it has
become possible to study
single quantum systems, such as trapped ions \cite{ions},
atoms in cavities \cite{atoms}, or even mesoscopically big
Cooper-pair boxes \cite{nakamura}. As a result, theoretical
consideration of the relation between dephasing and decoherence 
has become experimentally relevant and important.

In this work, we compare dephasing and decoherence in single
systems of interacting $s=1/2$ spins coupled to a bath of $s=1/2$ 
spins.
We show that in some cases, the oscillations can survive long
after the entropy has almost saturated, i.e. that the quantum
oscillations can take place for a long time even in an
almost completely decohered
system. These oscillations do not decay according to usual
exponential ($\exp{(-t/\tau)}$) or Gaussian ($\exp{(-t^2/\tau^2)}
$) law, but exhibit long power-law ($1/\sqrt{t}$ or $1/t$) tails.
This result has interesting consequences. 
The standard picture of a
decoherence process assumes that as soon as the system has
decayed into a mixture of pointer states, the fast quantum
mechanical motion is over, i.e.\ the pointer states are
essentially static. This has been confirmed by numerous studies
of different types of pointer states \cite{decgen}. 
However, we observe that
after the system has decayed into a mixture of the pointer
states, and its entropy has reached maximum, the oscillations
still persist. It means that the pointer states are not static:
they exhibit non-trivial dynamical behavior. We show this
explicitly by analyzing the structure of the density matrix.
The models considered here may be relevant for a number of
experimental systems \cite{stamprok}.

Studies of the dynamics of quantum oscillations have a long
history. In most situations considered so far, quantum oscillations
exhibit very fast exponential or Gaussian decay (it is the main
reason for the classical behavior of the world around us)
\cite{decgen}. Power-law-damped oscillations have been mentioned
\cite{stamprok} for a single spin decohered by a spin
bath, but the dynamics of decoherence and the structure of
pointer states has not been analyzed in detail. In contrast, in this
study the properties of the pointer states are of primary
importance, especially for a central system containing several
spins, where the non-trivial dynamics of the pointer states is
even more pronounced. Generally, to our knowledge, the
possibility of quantum oscillations without quantum coherence,
i.e.\ pointer states with non-trivial dynamics, have not been
discussed before.

For long-lasting oscillations, the system should be coupled
weakly to the environment, and the characteristic energies of the
system should be much larger than environmental ones. This
situation, being less relevant for experiments or the quantum
measurement problem, has not been studied in much detail.
Usually, in this case (referred to as the quantum limit of
decoherence \cite{qulim}), the pointer states are the eigenstates
$|\phi_n\rangle$ of the system's Hamiltonian. The decay of the
non-diagonal elements $\langle\phi_n|\rho |\phi_m\rangle$ of the
density matrix $\rho$ is Gaussian, and its rate is linearly
proportional to the magnitude of the interaction Hamiltonian
\cite{qulim}. However, as we show below, the quantum limit of
decoherence can be more subtle.

Analytical studies of dephasing/decoherence are not always
possible. In many cases, they include
approximations which can be quite stringent (e.g.\ Markovian
behavior of the bath). In this work, we solve directly
\cite{hans} the compound ``system-plus-bath'' time-dependent
Schr\"odinger equation, and the analytical approximations we use
are checked against this exact numerical solution.

First, let us study a single central spin $\bf s$
($s=1/2$) interacting with a bath of spins
${\bf I}_k$ ($I_k=1/2$), $k=1,\dots N$. The corresponding
Hamiltonian is
\begin{eqnarray}
{\cal H}={\cal H}_0+{\cal V}={\cal H}_S+{\cal H}_B+{\cal V},
\end{eqnarray}
where ${\cal H}_S$ and ${\cal H}_B$ are the Hamiltonians of
the system (central spin) and the bath, correspondingly,
and ${\cal V}$ is the system-bath
interaction. The central spin is subjected to an
external field $2\Delta$ applied along 
the $x$-axis, i.e.\ ${\cal H}_S = 2s_x \Delta = 
\sigma_x\Delta$ ($\sigma_{x,y,z}$ are Pauli's matrices).
We consider an Ising-type interaction
${\cal V}=\sigma_z \sum_{k=1}^N J_k \sigma^z_k$
(where $\sigma^z_k$ is Pauli's matrix of the spin ${\bf I}_k$),
and assume that the Hamiltonian of the bath
is zero, i.e.\ the bath has no internal dynamics.
Initially, the system and the bath are in an uncorrelated
product state $|s\rangle\otimes|b\rangle$; the state
of the bath $|b\rangle$ is a superposition
of all possible basis states with random coefficients.
This model can be used for description of decoherence
in various systems, from electron spins to SQUIDs
\cite{stamprok}; the
initial conditions then correspond to the temperature 
$\Delta\gg T\gg J_k$.

\begin{figure}
\includegraphics[width=8cm]{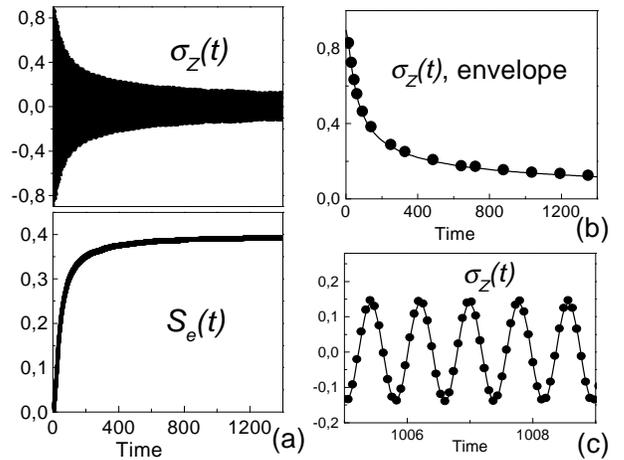}
\caption{Decoherence of a single spin coupled to a
static bath. (a): oscillations of $\sigma_z$ (top) and
an increase of entropy $S_e$ (bottom); the 
frequency of oscillations
is very high, and individual lines of the $\sigma_z(t)$
curve merge. (b) and (c): comparison of the analytical solution
(solid line)
with the numerical results (dots) for the envelope of 
$\sigma_z$ oscillations (b) and for oscillations themselves (c).
In (b), most of numerical data points have been
removed to make the line visible.}
\label{ising1}
\end{figure}

In Fig.\ \ref{ising1}, we show the results
for $N=14$ bath spins; the coupling
constants are randomly distributed from zero to
$J_k^{\text{max}}=0.125$:
$J_k=$\{0.123, 0.06425, 0.079, 0.009, 0.0585, 
0.03525, 0.012, 0.00525, 0.0945, 0.049, 0.1105, 0.02575, 
0.07625, 0.11225\}. 
Initial state of the central spin is defined by the
values $\sigma_x(0)=0.447$, $\sigma_y(0)=0$,
$\sigma_z(0)=0.894$; the external field $\Delta=4.0$. 
The observable $s_z(t)$ demonstrates damped 
oscillations: the central spin precesses around the
$x$-axis with gradually decreasing amplitude. 
Only $\sigma_z(t)$ is shown: oscillations of $\sigma_y$
are identical (only shifted in phase),
and $\sigma_x$ stays constant with good precision.
To characterize the dynamics of decoherence, we 
calculate the quadratic entropy
$S_e(t)=1-\mathop{\rm Tr}\rho^2(t)$ \cite{decgen},
Fig.\ \ref{ising1}(b).
It is more convenient than von Neumann's entropy,
while both characterize the same, how strongly mixed 
is the state of the system.

Initially the amplitude of oscillations drops,
and simultaneously the entropy rises very fast.
But, after the entropy has come close to its maximum,
and the system has decayed into a mixture of pointer states
$|p_1\rangle$ and $|p_2\rangle$,
the oscillations are clearly seen
for a very long time with the amplitude of about 20\% of the 
initial value. I.e., long-lasting quantum
oscillations can indeed exist in a decohered system. It means
that the pointer states
$|p_1\rangle$ and $|p_2\rangle$ have non-trivial dynamics.

Precision of the simulations can be checked since
this model is exactly solvable. The evolution operator is
$U(t) = \cos{\Omega t} - i(\sigma_z B + \sigma_x\Delta)
 \Omega^{-1} \sin{\Omega t}$,
where $\Omega = [\Delta^2+B^2]^{1/2}$, 
and $B = \sum J_k S^z_k$. Bath spins are static, and
the bath state $|b\rangle$
is a random superposition of a large number 
($2^{14}\approx 1.6\cdot 10^4$)
of the eigenstates of the operator $B$.
Therefore, by the central limit theorem, the
trace over the bath spins is equivalent to averaging over
the Gaussian random field $B$ with zero average 
and dispersion $b^2\approx\sum_k J_k^2$, and for 
the case $\Delta\gg b$:
\begin{eqnarray}
\label{szt}
\sigma_z(t)&=&\frac{\sigma_z(0)}{2\pi b^2}
  \mathop{\rm Re} \int dB \exp{\left[-2 i\Delta t
  -iB^2t/\Delta\right]}\\
 \nonumber
  &\times &\exp{[-(1/2)B^2/b^2]}
\end{eqnarray}
The envelope of the $\sigma_z(t)$ oscillations is 
$\sigma^{\text{(env)}}_z(t)=\sigma_z(0)%
\left[1+(2t/\tau_1)^2\right]^{1/4}$, where
$\tau_1=\Delta/b^2$.
Thus, initially there is the usual Gaussian damping
(quadratic with time), but afterwards it changes to
a slow power-law decay $1/\sqrt{2t/\tau_1}$.
The analytical results for the envelope, and for the
oscillations of $\sigma_z(t)$ themselves, coincide
perfectly with the numerical simulations, see Fig.\
\ref{ising1}(c) and (d).

The analytical solution makes clear that the rapidness of 
the system's
dynamics ($\Delta\gg J_k$ in our case) is
needed for long-time oscillations in a
decohered system. Fast motion of the system eliminates
from the evolution operator the decohering terms which
are of first order in the system-bath coupling $B$, and
only second-order terms survive, as in Eq.\ \ref{szt}. 
In contrast to the conclusion of Ref.\ 
\cite{qulim}, the decoherence remains rather fast, although
the spin-bath interaction does not have diagonal elements
in the $\sigma_z$ basis.

\begin{figure}
\includegraphics[width=8cm]{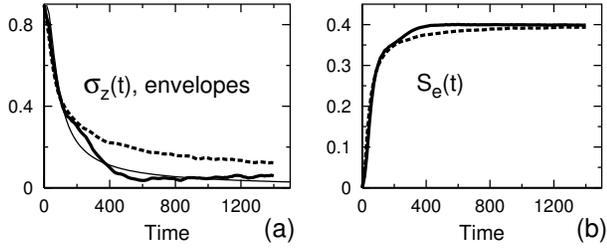}
\caption{Decoherence of a single spin coupled to a
slow bath. (a): the envelopes of $\sigma_z(t)$ oscillations,
(b): entropy. Thick solid line --- numerical
results for $h_x$ varying from 0.005 to 1.0 (all lines
almost coincide), dashed line --- the case of $h_x=0$,
thin solid line --- analytical solution.}
\label{isinghx}
\end{figure}

We have considered above a static 
environment, but slow internal dynamics does
not influence our qualitative conclusions. 
As an example, we add an external field
acting on the bath spins, so the Hamiltonian
of the bath becomes ${\cal H}_B = h_x \sum_k S_k^x$,
where $h_x\ll \Delta$ (all other parameters are
kept the same). The results of simulations
are shown in Fig.\ \ref{isinghx}.
Again, the slowly decaying oscillations are seen 
for long times, long after the entropy came almost to
saturation, but the oscillations decay 
faster than in the case of zero $h_x$. 
Rather interestingly, the decay of the $\sigma_z(t)$
oscillations is practically independent
of the $h_x$ value for $h_x$ varying from 0.005 to 1.0
(i.e.\ by more than two orders of magnitude).

An exact analytical solution is not available for $h_x\neq 0$.
To analyze the dynamics of the system, we use the 
cumulant expansion 
for the evolution operator
(the Magnus expansion, widely used e.g.\
in the theory of NMR)\cite{waugh}:
$U(t)=\exp{(-i{\cal H}_0 t)}\exp{[-iX(t)]}$, where
\begin{equation}
X(t)=\int_0^t ds {\cal V}(s) - 
  \frac{i}{2}\int_0^t\! \int_0^s\!ds\, du\,
  [{\cal V}(s),{\cal V}(u)]+\dots
\end{equation}
where ${\cal V}(t)=\exp{(i{\cal H}_0 t)}{\cal V}
\exp{(-i{\cal H}_0 t)}$, and only secular terms
(growing linearly with time without oscillations)
are retained. 
At short times $t\ll \tau_2=1/h_x$, the bath is
practically static, and the results are
the same as for the case $h_x=0$.
For long times $t\gg \tau_2$, independently of the value of 
$h_x$,
\begin{equation}
\label{cumU1}
U(t)= \exp{\left[-i\sigma_x t \Delta 
  -i\sigma_x t (B_z^2+B_y^2)/(4\Delta)\right]},
\end{equation}
where $B_z=\sum_k J_k S_k^z$, and $B_y=\sum_k J_k S_k^y$.
To evaluate the evolution operator (\ref{cumU1}), we
consider the bath in the mean-field manner
similar to the Mermin model
\cite{mermin}, and replace $B_{y,z}$ by Gaussian random fields.
The envelope is
$\sigma^{\text{(env)}}_z(t)=\sigma_z(0)\left%
[1+(b^2t/\Delta)^2\right]^{1/2}$, i.e. the oscillations
decay as $1/t$.
These conclusions agree reasonably with the results of
numerical simulations, see Fig.\ \ref{isinghx}(a):
for $h_x$ varying from $5\cdot 10^{-3}$ to 1.0,
the shape of the envelope of $\sigma_z(t)$ does not
depend on $h_x$ and is close to the analytical curve.

Another way of turning on the dynamics of the bath
is to couple bath spins with each other by small exchange
interactions. We have studied the Ising-type exchange 
${\cal H}_B=\sum_k A_{kl} S^x_k S^x_l$, with constant
$A_{kl}=A$ and random $A_{kl}$. The 
results are qualitatively the same as above, what
agrees with our considerations and arguments
of Ref.\ \cite{qulim}, which suggest that the 
exact dynamics of the bath
is unimportant as long as it is slow.

Finally, we study the quantum oscillations
in decohered many-spin systems. In this case, we see
most clearly the two stages, the first one associated with
the decay of the system into a mixture of dynamical
pointer states (when the entropy rises close to its 
maximum), and the second one, when the oscillations of the
pointer states decay.

\begin{figure}
\includegraphics[width=8cm]{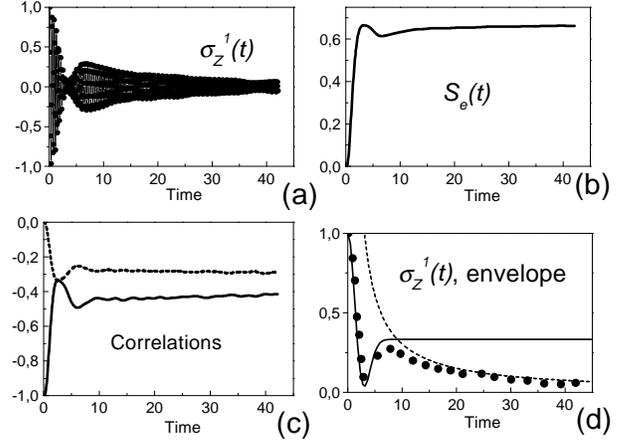}
\caption{Decoherence of two coupled spins by a spin
bath. (a): oscillations of $\sigma_z^1(t)$; (b):
entropy; (c): correlations between the central
spins, $C_{12}^{zz}$ (solid line), and
$C_{12}^{xx}=C_{12}^{yy}$ (dashed line); (d): comparison of the
numerical results for $\sigma_z^1(t)$ (dots) with 
analytical short-time solution (solid line) and long-time
approximation (dashed line). Most of numerical data points
are removed to make the lines visible.}
\label{heis1}
\end{figure}

As a simple example,
we consider two spins ${\bf s}_1$ and ${\bf s}_2$ 
with Heisenberg anisotropic coupling $J$ between them,
so that the system's Hamiltonian 
${\cal H}_0=2J{\bf s}_1{\bf s}_2+J/2$. Spins are coupled
to the static bath via isotropic Heisenberg exchange, 
${\cal V}=\sum_k J_k ({\bf s}_1+{\bf s}_2) {\bf S}_k$.
For the simulations presented in Fig.\ \ref{heis1}, 
we use $J=8.0$,
and the values of the coupling constants are the same
as above. The initial state of the system is the product
$|\uparrow\rangle |\downarrow\rangle$, i.e.\ the symmetric
superposition of the triplet 
$|s=1, s_z=0\rangle$ and the singlet $|s=0\rangle$ states.
We present only the oscillations
of the $z$-component of the first spin $\sigma_1^z(t)$:
oscillations of 
$\sigma_2^z(t)$ are just shifted in phase
by $\pi$, and all other components of the central spins
remain practically constant. 

The results of the numerical simulations are
shown in Fig.\ \ref{heis1} (note the time scale
in this case). Again, there is an initial sharp
decrease of the amplitude of oscillations and an increase
of entropy; moreover, the correlations
$C_{12}^{\alpha\beta} =\langle\sigma_1^\alpha \sigma_2^\beta\rangle$
between the
two central spins follow the behavior of the entropy
supporting the conclusion about practically complete
initial decoherence of the system. Nevertheless,
again, the oscillations
exhibit a long tail, although
the entropy and the correlations has saturated. 
Analytical treatment of the Heisenberg case also supports
this picture. The exact evolution operator is
\begin{equation}
\label{heisU}
U(t)=\exp{[-iJ{\bf s}^2 t/2-i{\bf s}\sum_k J_k {\bf S}_k]},
\end{equation}
and its mean-field evaluation gives 
$\sigma_1^{z\text{(env)}}(t)=(1/3)[1-2(b^2t^2-1)\exp{(-b^2t^2/2)}]$,
i.e.\ usual
fast Gaussian decay, but the final value is 1/3 rather
than zero. This prediction, valid for short times, is
in very good agreement with the numerical results,
see Fig.\ \ref{heis1} (d). However, there is a 
subsequent decay of oscillations, 
associated with the next-order terms, quadratic in
$b$. We did not manage to obtain an analytical expression, but
in analogy with the single-spin $h_x\neq 0$ case, we
can assume $1/t$ decay, and indeed, the numerical
results in Fig.\ \ref{heis1}(d)
are in good agreement with the $1/t$ assumption.

\begin{figure}
\includegraphics[width=8cm]{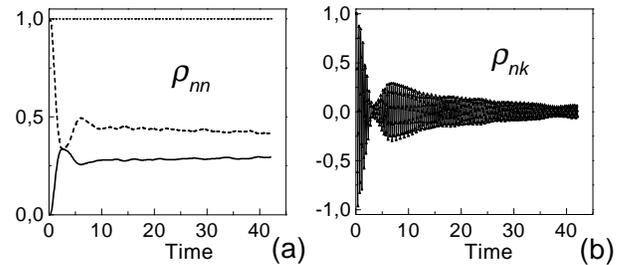}
\caption{Dependence of the density matrix elements vs.\ time.
(a): diagonal elements for $|s=1,s_z=0\rangle$ state
(solid line), $|s=1,s_z=\pm 1\rangle$ states (dashed line),
 and $|s=0\rangle$ state (dotted line); (b): nondiagonal
 element $\langle s=0|\rho|s=1,s_z=0\rangle$. All other
 matrix elements are very small.}
\label{heis2}
\end{figure}

Detailed analysis of correlations between the two
spins allows us to reconstruct the full density matrix,
and to check the structure of pointer states explicitly.
The results are shown in Fig.\ \ref{heis2}. 
At the first stage, the diagonal
elements of the density matrix change: 
part of the $|s=1,s_z=0\rangle$ spectral weight
is transferred equally to the $|s=1,s_z=\pm 1\rangle$ 
states. The direction of the system's 
total spin randomizes (although incompletely) due to 
rotation around the randomly oriented effective field
generated by the bath
(see Eq.\ \ref{heisU}). Also, at the first
stage the non-diagonal element of the density
matrix $\langle s=0|\rho|s=1,s_z=0\rangle$ decays rapidly,
not to zero, but to 0.3. The other non-diagonal elements
remain small (not shown). Thus, by the end of the first
stage, the system has decayed into a mixture of
pointer states belonging to the subspaces of different
$s_z$. However, these pointer states are not static:
there are long-living oscillations within the subspace $s_z=0$
(containing two states, one singlet and one triplet).
These oscillations do not take place
inside decoherence-free subspaces (DFS) \cite{dfs,dfs1},
and are due to non-trival fast quantum dynamics of the
pointer states. This picture holds also for other initial
conditions, but, in general, the dynamics inside the DFS and the
oscillations of pointer states show up simultaneously.

The conclusions presented here are confirmed (results not shown)
by calculations for different sets of the system's parameters,
for the environments of different sizes, etc. We have also
checked the case when the central spins are coupled differently 
with the environment, and the picture remains the
same. Our qualitative conclusions also hold for larger
central systems, such as a ring of 4 coupled spins.

In summary, we have studied numerically the dephasing and
decoherence processes which take place in some generic systems of
interacting spins 1/2 coupled to a spin bath. We compare the
dynamical increase of entropy of the central system and the decay
of quantum oscillations. We found that in some cases quantum
oscillations take place long after the entropy came close to
saturation, i.e.\ that the quantum coherent oscillations can
exist in decohered systems. The oscillations exhibit long tails
which decay with time as $1/t$ or $1/\sqrt{t}$, and are
observable long after the system has decayed into a mixture of
pointer states. Therefore, the pointer states can exhibit non-
trivial fast dynamics. We have shown this also by direct
analysis of the density matrix. 

The authors would like to thank A. Melikidze for helpful discussions.
This work was partially carried out at the Ames Laboratory, which
is operated for the U.\ S.\ Department of Energy by Iowa State
University under Contract No.\ W-7405-82 and was supported by the
Director of the Office of Science, Office of Basic Energy
Research of the U.\ S.\ Department of Energy. Support from the
Dutch ``Stichting Nationale Computer Faciliteiten (NCF)''
is gratefully acknowledged. This work was partially supported by
Russian Basic Research Foundation, grant 01-02-16108.

\end{document}